\documentclass[a4paper]{jpconf}
\usepackage{graphicx}

\usepackage{hyperref}
\usepackage[all]{hypcap}
\hypersetup{
    colorlinks,
    citecolor=blue,
    filecolor=blue,
    linkcolor=black,
    urlcolor=blue
}
\begin{document}
\def\bc{\begin{center}}
\def\ec{\end{center}}
\def\beq{\begin{equation}}
\def\eeq{\end{equation}}
\def\noi{\noindent}
\def\hs#1{\hspace*{#1cm}}
\def\cov#1#2{{\textrm{cov}(#1,#2)}}

\def\av#1{\langle #1 \rangle}
\def\avcond#1{\langle #1 \!\rangle}
\def\avav#1{\langle\langle #1 \rangle\rangle}
\def\avL#1{\left\langle #1 \right\rangle}
\def\avo#1{{\av{#1}}}
\def\avr#1#2{\langle {#1} \rangle^{}_{#2}}
\def\Av#1{{\left\langle #1 \right\rangle}}
\def\Avr#1#2{\left\langle {#1} \right\rangle_{#2}}
\def\Avup#1#2{\left\langle {#1} \right\rangle^{#2}}
\def\AvC#1#2#3{\left\langle {#1} \right\rangle^{#2}_{#3}}

\def\pir{{\frac{1}{2\pi}}}
\def\pirr{{(2\pi)^{-2}_{}}}

\def\nF{N_F}
\def\nB{N_B}
\def\nuF{\nu_F}
\def\nuB{\nu_B}
\def\nuPF{\nu_{P_F}}
\def\nuPB{\nu_{P_B}}
\def\gamF{\gamma_F}
\def\gamB{\gamma_B}

\def\nFr{{\frac{\nF}{\av{\nF}}}}
\def\nBr{{\frac{\nB}{\av{\nB}}}}
\def\pF{p_{TF}}
\def\pB{p_{TB}}
\def\PF{P_{TF}}
\def\PB{P_{TB}}
\def\mF{{\mu^{}_F}}
\def\mB{{\mu^{}_B}}
\def\mFr{{\frac{\mF}{\av{\mF}}}}
\def\mBr{{\frac{\mB}{\av{\mB}}}}

\def\omu{{\av{\mu}}}
\def\omF{{\av{\mu^{}_F}}}
\def\omB{{\av{\mu^{}_B}}}
\def\omBF{{\av{\mu^{}_F\mu^{}_B}}}
\def\omFB{{\av{\mu^{}_F\mu^{}_B}}}

\def\mo{{\mu^{}_{0}}}
\def\moo{{\mu^{2}_{0}}}
\def\rhoo{{\rho^{}_{0}}}
\def\rhooo{{\rho^{2}_{0}}}
\def\moF{{\mu^{}_{0F}}}
\def\moB{{\mu^{}_{0B}}}

\def\ommF{{\av{\mu_F^2}}}
\def\ommB{{\av{\mu_B^2}}}
\def\omFF{{\av{\mu_F^{}}_{}^2}}
\def\omBB{{\av{\mu_B^{}}_{}^2}}
\def\df{\delta_{F,\sum F_i}}
\def\db{\delta_{B,\sum B_i}}
\def\pp{\prod_{i=1}^N p(B_i,F_i)}
\def\sumn{\sum^n_{i=1}}
\def\sumNstr{\sum^{\infty}_{N_{str}=0}P\left(N_{str}\right)}
\def\sumnB{\sum^{\infty}_{n_{B}=0}}
\def\sumnF{\sum^{\infty}_{n_{F}=0}}
\def\sumnBOne{\sum^{\infty}_{n_{B_{1}}=0}}
\def\sumnFOne{\sum^{\infty}_{n_{F_{1}}=0}}
\def\sumnBNstr{\sum^{\infty}_{n_{B_{N_{str}}}=0}}
\def\sumnFNstr{\sum^{\infty}_{n_{F_{N_{str}}}=0}}

\def\oN{\overline{N}}
\def\obr{\overline{b}^{rel}_{}}
\def\brel{b_{rel}^{}}
\def\bm{b_{mod}^{}}
\def\brob{b_{rob}^{}}
\def\babs{b_{abs}^{}}
\def\ba{b_{abs}^{}}
\def\bsym{b_{sym}^{}}
\def\brob{b_{rob}^{}}
\def\oba{\overline{b}^{abs}_{}}
\def\ar{a^{rel}_{}}
\def\aa{a^{abs}_{}}
\def\dnF{\nF-\av{\nF}}
\def\pc{\!+\!}
\def\mc{\!-\!}
\def\ppc{\!+\!...\!+\!}
\def\yFB{{\eta^{}_{sep}}}
\def\yBF{{\eta^{}_{sep}}}
\def\yF{{\eta^{}_F}}
\def\yB{{\eta^{}_B}}
\def\fFB{{\phi^{}_{sep}}}
\def\fBF{{\phi^{}_{sep}}}
\def\fF{{\phi^{}_F}}
\def\fB{{\phi^{}_B}}

\def\dyf{{d\eta d\phi}}
\def\dyfo{{d\eta_1 d\phi_1}}
\def\dyft{{d\eta_2 d\phi_2}}
\def\dyp{{d\eta'}}
\def\dfp{{d\phi'}}
\def\dyfp{{d\eta' d\phi'}}

\def\Dyy{{(\delta \eta)^2}}
\def\Dy{{\delta \eta}}
\def\Df{{\delta \phi}}

\def\DyF{{\delta \eta^{}_F}}
\def\DyFF{{(\delta \eta^{}_F)^2}}
\def\DyB{{\delta \eta^{}_B}}

\def\DfFF{{(\delta \phi^{}_F)^2}}
\def\DfF{{\delta \phi^{}_F}}
\def\DfB{{\delta \phi^{}_B}}

\def\DyfFF{{(\delta \eta^{}_F\delta \phi^{}_F)^2}}
\def\DyfF{{\delta \eta^{}_F\delta \phi^{}_F}}
\def\DyfB{{\delta \eta^{}_B\delta \phi^{}_B}}

\def\aFF{{\delta^{2}_F}}
\def\aFFr{{\delta^{-2}_F}}
\def\aF{{\delta^{}_F}}
\def\aB{{\delta^{}_B}}
\def\aFr{{\delta^{-1}_F}}
\def\aBr{{\delta^{-1}_B}}
\def\aW{{\delta^{}_W}}

\def\acF{{\delta \eta^{}_F\delta \phi^{}_F/2\pi}}
\def\acB{{\delta \eta^{}_B\delta \phi^{}_B/2\pi}}
\def\acW{{\delta \eta\delta \phi/2\pi}}

\def\eg{{\eta^{}_{gap}}}
\def\yg{{\eta^{}_{gap}}}
\def\fg{{\phi^{}_{gap}}}

\def\omN{{\omega_N}}

\def\bp{\textbf{p}}
\def\oq{\overline{q}}
\def\fv{\phi}
\def\loy{\lambda_1(\eta)}
\def\loyo{\lambda_1(\eta_1)}
\def\loyt{\lambda_1(\eta_2)}
\def\lty{\lambda_2(\eta_1,\eta_2)}
\def\lo#1{\lambda_1(#1)}
\def\loo#1{\lambda^2_1(#1)}
\def\lt#1{\lambda_2(#1)}
\def\Lam#1{\Lambda(#1)}
\def\tLam#1{\widetilde{\Lambda}(#1)}

\def\loyf{\lambda_1(\eta,\phi)}
\def\loyfo{\lambda_1(\eta_1,\phi_1)}
\def\loyft{\lambda_1(\eta_2,\phi_2)}
\def\ltyf{\lambda_2(\eta_1,\phi_1;\eta_2,\phi_2)}

\def\roy{\rho_1(\eta)}
\def\royo{\rho_1(\eta_1)}
\def\royt{\rho_1(\eta_2)}
\def\rty{\rho_2(\eta_1,\eta_2)}
\def\ro#1{\rho_1(#1)}
\def\roo#1{\rho^2_1(#1)}
\def\rt#1{\rho_2(#1)}
\def\royf{\rho_1(\eta,\phi)}
\def\royfo{\rho_1(\eta_1,\phi_1)}
\def\royft{\rho_1(\eta_2,\phi_2)}
\def\rtyf{\rho_2(\eta_1,\phi_1;\eta_2,\phi_2)}

\def\IFF{{I_{F\!F}^{}}}
\def\IBF{{I_{F\!B}^{}}}
\def\IBB{{I_{B\!B}^{}}}
\def\JFF{{J_{F\!F}^{}}}
\def\JBF{{J_{F\!B}^{}}}
\def\sigFB{{\sigma^2_{\nF+\nB}}}
\def\sigF{{\sigma^2_{\nF}}}
\def\sigB{{\sigma^2_{\nB}}}
\def\sigN{{\sigma^2_N}}
\def\pT{{p_{\rm T}}}
\def\pT{{p_{\rm T}}}
\def\etaF{{\eta_F}}
\def\etaB{{\eta_B}}
\def\phiF{{\phi_F}}
\def\phiB{{\phi_B}}
\def\detaF{{\delta\eta_F}}
\def\detaB{{\delta\eta_B}}
\def\dphiF{{\delta\phi_F}}
\def\dphiB{{\delta\phi_B}}
\def\pBj{{\pB^j}}
\def\pFi{{\pF^i}}
\def\avpF{{\avav c_F}}

\def\avpF{{\langle\langle p_t \rangle\rangle_F^{}}}
\def\avpB{{\langle\langle p_t \rangle\rangle_B^{}}}
\def\avpFpF{{\langle\langle p^{2}_t \rangle\rangle_F^{}}}
\def\avpBpB{{\langle\langle p^{2}_t \rangle\rangle_B^{}}}
\def\avpBpBtwoparticle{{\langle\langle p_t,p_t \rangle\rangle_B^{}}}
\def\avpFpFtwoparticle{{\langle\langle p_t,p_t \rangle\rangle_F^{}}}
\def\avpBpFtwoparticle{{\langle\langle p_t,p_t \rangle\rangle_{B,F}^{}}}

\def\bpp{b_{p_tp_t}}
\def\bpn{b_{p_tn}}
\def\bnn{b_{nn}}

\def\pp{p_t\!-\!p_t}
\def\pn{p_t\!-\!n}
\def\nn{n\!-\!n}
\def\DpDp{\Delta p_t\!-\!\Delta p_t}

\def\aDpt{{\av{\Delta p_t\Delta p_t}}}
\def\Dpt{{\av{\Delta p_t\Delta p_t}(\etaF,\phiF,\etaB,\phiB)}}

\def\ccdot{\!\cdot\!}

\title{$N-N$, $P_{T}-N$ and $P_{T}-P_{T}$ fluctuations in nucleus-nucleus collisions at the NA61/SHINE experiment}

\author{Evgeny Andronov for the NA61/SHINE Collaboration}

\address{Saint Petersburg State University, ul. Ulyanovskaya 1, 198504, Petrodvorets, Saint Petersburg, Russia}

\ead{e.v.andronov@spbu.ru}

\begin{abstract}
The NA61/SHINE experiment aims to discover the critical point of strongly interacting matter and study the properties of the onset of deconfinement. For these goals a scan of the two dimensional phase diagram ($T-\mu_{B}$) is being performed at the SPS by measurements of hadron production in proton-proton, proton-nucleus and nucleus-nucleus interactions as a function of collision energy. This paper presents preliminary results from Be+Be collisions on pseudorapidity dependences of transverse momentum and multiplicity fluctuations expressed in terms of strongly intensive quantities. It is shown that non-trivial effects evolve from the Poissonian-like fluctuations for small pseudorapidity intervals with expansion of the acceptance. These fluctuations are supposed to be sensitive to the existence of the critical point. The results will be compared to the predictions from the EPOS model.
\end{abstract}

\section{Introduction}
The NA61/SHINE experiment~\cite{Abgrall:2014fa} is a multi-purpose fixed target experiment at the Super Proton Synchrotron (SPS) of the European Organization for Nuclear Research (CERN). The strong interactions programme of NA61/SHINE consists of studies of the onset of deconfinement (OD)~\cite{Gazdzicki:1998vd} in nucleus-nucleus collisions and search for the critical point (CP)~\cite{Fodor:2004nz} of strongly interacting matter.

NA61/SHINE performs measurements of hadron production in collisions of protons and various nuclei (p+p, p+Pb, Be+Be, Ar+Sc, Xe+La, Pb+Pb) in a range of beam momenta (13{\it A} - 150/158{\it A} GeV/c). It is expected that there will be a non-monotonic dependence of fluctuations of a number of observables on energy and system size in this scan due to the phase transition of strongly interacting matter and the possible existence of the CP~\cite{Stephanov:1999zu}. Some hints of such behaviour were already observed by the NA49 experiment~\cite{Grebieszkow:2009jr}.
\section{$P_{T}-N$ fluctuation measures}
The volume of the system created in ultrarelativistic ion collisions fluctuates from event to event. In order to suppress contribution from this "trivial" fluctuations strongly intensive observables are used \cite{Gorenstein:2011vq, Gazdzicki:2013ana}: 
\begin{eqnarray}
            &\Delta[A,B] = \frac{1}{C_{\Delta}} \biggl[ \langle B \rangle \omega[A] -
                        \langle A \rangle \omega[B] \biggr] \\
            &\Sigma[A,B] = \frac{1}{C_{\Sigma}} \biggl[ \langle B \rangle \omega[A] +
                        \langle A \rangle \omega[B] - 2 \bigl( \langle AB \rangle -
                        \langle A \rangle \langle B \rangle \bigr) \biggr],
\end{eqnarray}
where $\langle\cdots\rangle$ stands for averaging over all events and $\omega[X]$ is the scaled variance of any quantity $X$ defined as $(\langle X^{2}\rangle -{\langle X\rangle}^{2})/\langle X\rangle$. In case of joint fluctuations of the total transverse momentum $P_{T}$ and multiplicity $N$ one defines:
$A=P_{T}=\sum_{i=1}^{N}p_{T_{i}}$, $B=N$, $C_{\Delta}=C_{\Sigma}=\langle N\rangle \omega[p_{T}]$, where $\omega[p_{T}]$ is the scaled variance of the inclusive $p_{T}$ spectrum.

The advantage of the strongly intensive quantities is that in the models of independent sources \cite{Bialas:1976ed} and in the models of independent particle production the contribution from volume fluctuations is eliminated, allowing to probe the genuine CP signals. Moreover, in the models of independent particle production $\Delta[P_{T},N]=\Sigma[P_{T},N]=1$. These equalities also hold for the ideal Boltzmann gas both in the grand canonical and in the canonical ensemble formulations. Another trivial property of these quantities is that they vanish in case of absence of fluctuations of $A$ and $B$. 

In recent measurements by the NA61/SHINE collaboration \cite{Czopowicz,Andronov:cpod} of the strongly intensive quantities $\Delta[P_{T},N]$ and $\Sigma[P_{T},N]$ no anomaly attributable to the CP was observed neither in p+p nor in forward energy selected Be+Be and Ar+Sc collisions. This observation led to the extension of the analysis to studies of the pseudorapidity dependence of strongly intensive observables. By changing the width of the chosen pseudorapidity interval one can probe different values of baryochemical potential as evident from the different shapes of proton and antiproton rapidity spectra~\cite{Pulawski}. Therefore, the new analysis allows to probe additional regions of the phase diagram of strongly interacting matter.

In this paper new results were obtained for Be+Be collisions at 150{\it A} GeV/c selected for the smallest 8$\%$ of forward energies. This selection of the most central events by the forward energy was done using information from the NA61/SHINE forward calorimeter PSD. This detector consists of 44 square modules. A specific set of modules is selected in order to maximize the signal from projectile nucleon spectators and to suppress background from produced particles. Analysis was performed for the 120k selected events containing 2000k quality selected tracks in total.

Fluctuations were studied in 9 pseudorapidity intervals defined in the laboratory reference frame using all charged particles produced in strong and electromagnetic processes from the primary vertex. The forward edge of the intervals was fixed at 5.2 units of pseudorapidity, with the backward edge changing from 3 up to 4.6 units. The choice of lower bounds was motivated by the small azimuthal angle acceptance at smaller pseudorapidities. The upper bound was introduced in order to suppress possible quasi-diffractive or electromagnetic effects which become important at larger pseudorapidities. Transverse momenta of all charged particles were restricted to $0<p_{T}<1.5$ GeV/c. Moreover, the NA61/SHINE acceptance map \cite{Acceptance} was applied.

\begin{figure}[h]
\begin{minipage}{18pc}
\includegraphics[width=18pc]{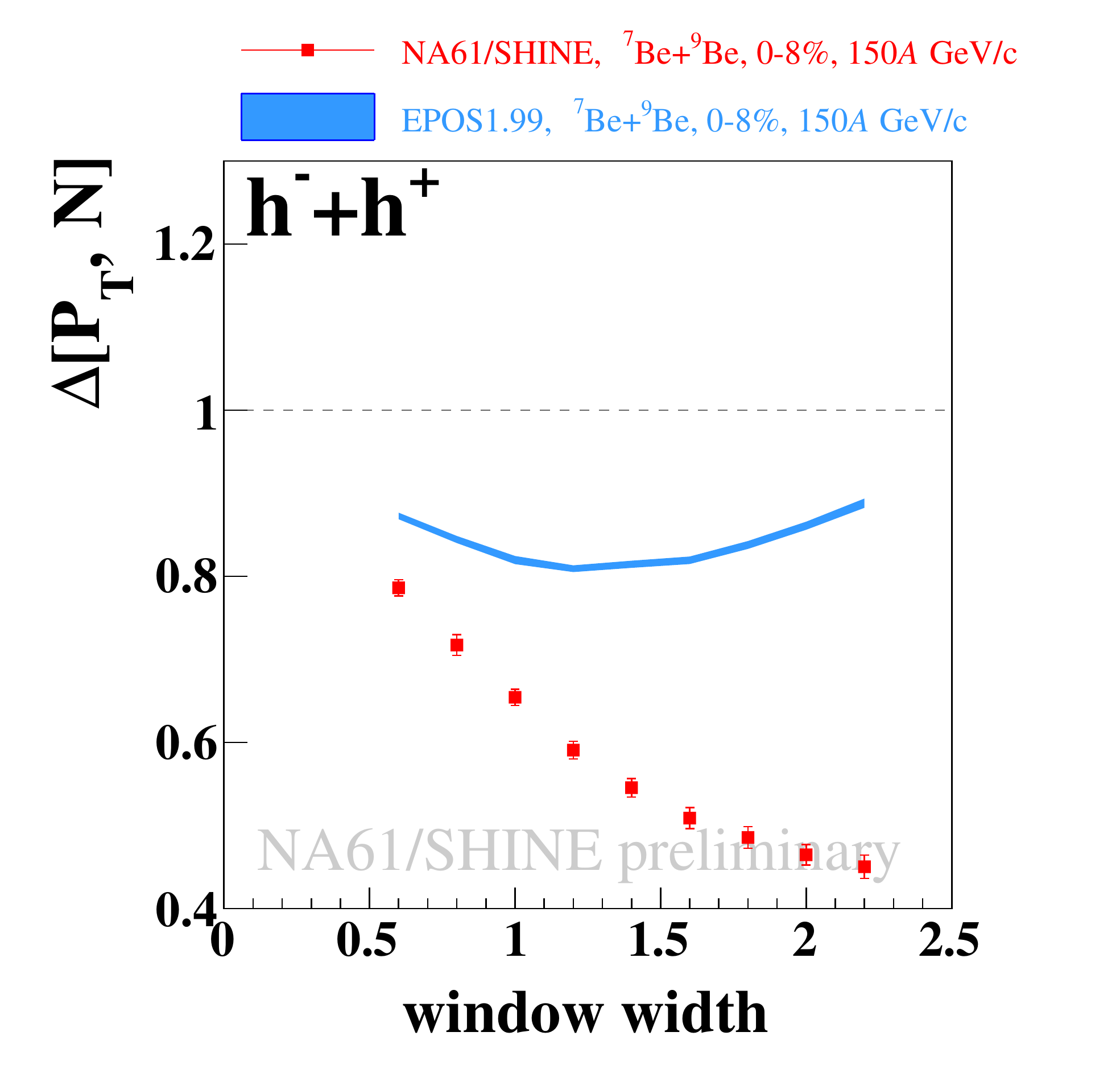} \begin{center}(a)\end{center}
\end{minipage}
\begin{minipage}{18pc}
\includegraphics[width=18pc]{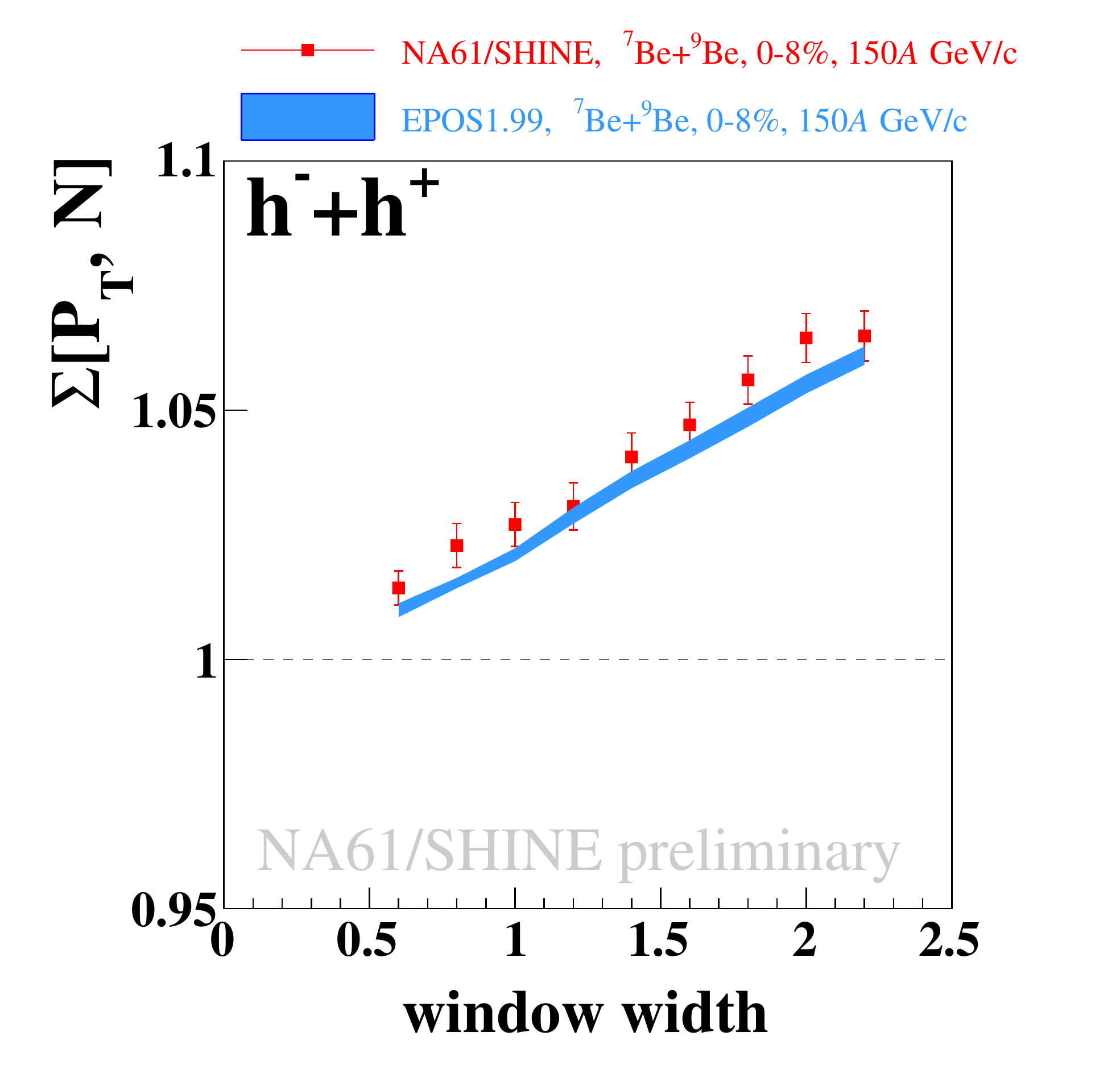} \begin{center}(b)\end{center}
\end{minipage}
\caption{\label{label1}Dependence of $\Delta[P_{T},N]$ (a) and $\Sigma[P_{T},N]$ (b) on the width of the pseudorapidity window. Red squares are preliminary NA61/SHINE measurements for 0-8$\%$ Be+Be collisions at 150{\it A} GeV/c. Blue band represents the EPOS1.99 model predictions.}
\end{figure}

Figure \ref{label1} shows preliminary results on the dependence on the width of the pseudorapidity interval for $\Delta[P_{T},N]$ and $\Sigma[P_{T},N]$ and comparisons to the EPOS1.99 model \cite{Pierog:2009zt} predictions. Both quantities change monotonously for the data in contrast to the EPOS1.99 results for $\Delta[P_{T},N]$ which show a minimum for intermediate width and lie significantly above the measurements. EPOS1.99 also fails to describe $\Delta[P_{T},N]$ for the full NA61/SHINE pseudorapidity acceptance in inelastic p+p interactions \cite{Czopowicz}. Another observation is that for smaller windows these quantities approach unity (independent particle production limit) as the number of particles in the interval gets small. Moreover, the inequalities $\Delta[P_{T},N]<1$ and $\Sigma[P_{T},N]\geq 1$, previously observed for all studied systems at all collision energies \cite{Czopowicz,Andronov:cpod}, also hold when modifying the width of the pseudorapidity window of the measurement.  In general, no traces of the possible critical point of strongly interacting matter are visible.  

The results were not corrected for detector inefficiencies and trigger biases as simulations have shown that their effect estimated using the GEANT3 package does not exceed 5$\%$. In contrast the trigger bias modifies results significantly for p+p collisions~\cite{Czopowicz}.

The statistical uncertainties were determined using the sub-sample method. The systematic uncertainties were estimated to be smaller than 10$\%$ for $\Delta[P_{T},N]$ and $\Sigma[P_{T},N]$. Further analysis of the systematic uncertainties will follow.

\section{$N-N$, $P_{T}-N$ and $P_{T}-P_{T}$ fluctuations in two rapidity intervals}

By analogy with the $P_{T}-N$ case described in the previous section one can introduce a number of other strongly intensive observables. Joint fluctuations of two quantities in two separated pseudorapidity intervals are of special interest because they are closely connected with studies of forward-backward correlations that have a long history of measurements \cite{Corr:ee, Corr:Alice}. Forward-backward correlations are usually quantified by the correlation coefficient which is not strongly intensive and is sensitive to the employed centrality selection procedures in nucleus-nucleus collisions \cite{Corr:STAR}. Use of strongly intensive quantities can suppress these trivial fluctuations and allows to study the intrinsic properties of particle emitting sources. Therefore, for studies of joint fluctuations of multiplicities ($N_{F}$, $N_{B}$) and the sum of transverse momenta ($P_{TF}$, $P_{TB}$) in two windows of rapidity the following quantities are defined~\cite{Andronov:sigma}: 

\beq\label{sigmaNBF}
\Sigma[\nF,\nB] = \frac{1}{\av \nB +\av \nF}\cdot \biggl[ \av \nB  \omega[\nF] +
                        \av \nF  \omega[\nB] - 2 \bigl( \av {\nF\nB}  -
                        \av \nF  \av \nB  \bigr)\biggr]
\eeq
$$
\Sigma[\nF,\PB] = \frac{1}{(\av \nB +\av \nF)\cdot\overline{p_{T}}_{B} +\av\nF\omega[p_{T}]_{B}}\cdot
$$
\beq\label{sigmaPTNBF}
 \biggl[ \av \PB  \omega[\nF] +
                        \av \nF  \omega[\PB] - 2 \bigl( \av {\nF\PB}  -
                        \av \nF  \av \PB  \bigr)\biggr]
\eeq
$$
\Sigma[\PF,\PB] = \frac{1}{\av\PB\left(\overline{p_{T}}_{F}+\omega[p_{T}]_{F}\right) +\av\PF\left(\overline{p_{T}}_{B}+\omega[p_{T}]_{B}\right)}\cdot 
$$
\beq\label{sigmaPTPTBF}
 \biggl[ \av \PB  \omega[\PF] +
                        \av \PF  \omega[\PB] - 2 \bigl( \av {\PF\PB}  -
                        \av \PF  \av \PB  \bigr)\biggr]
\eeq
Here, averaging over all tracks in the forward (backward) window is denoted as:
\beq\label{av-tr}
\overline{p_{T}}_{F,B}=\frac{1}{N_{tr}}\sum_{tr_{F,B}} p_{T}
  \ .
\eeq
Moreover, the scaled variance of inclusive $p_{T}$ spectra $\omega[p_{T}]_{F,B}$ is introduced for both forward and backward windows.

Typically, forward-backward correlations are studied not for the sum of transverse momenta but for the event mean transverse momentum \cite{Braun}, but in order to construct a strongly intensive quantity from two observables $A$ and $B$ both of these quantities have to be extensive \cite{Gorenstein:2011vq}. 

\begin{figure}[h]
\begin{minipage}{12pc}
\includegraphics[width=12pc]{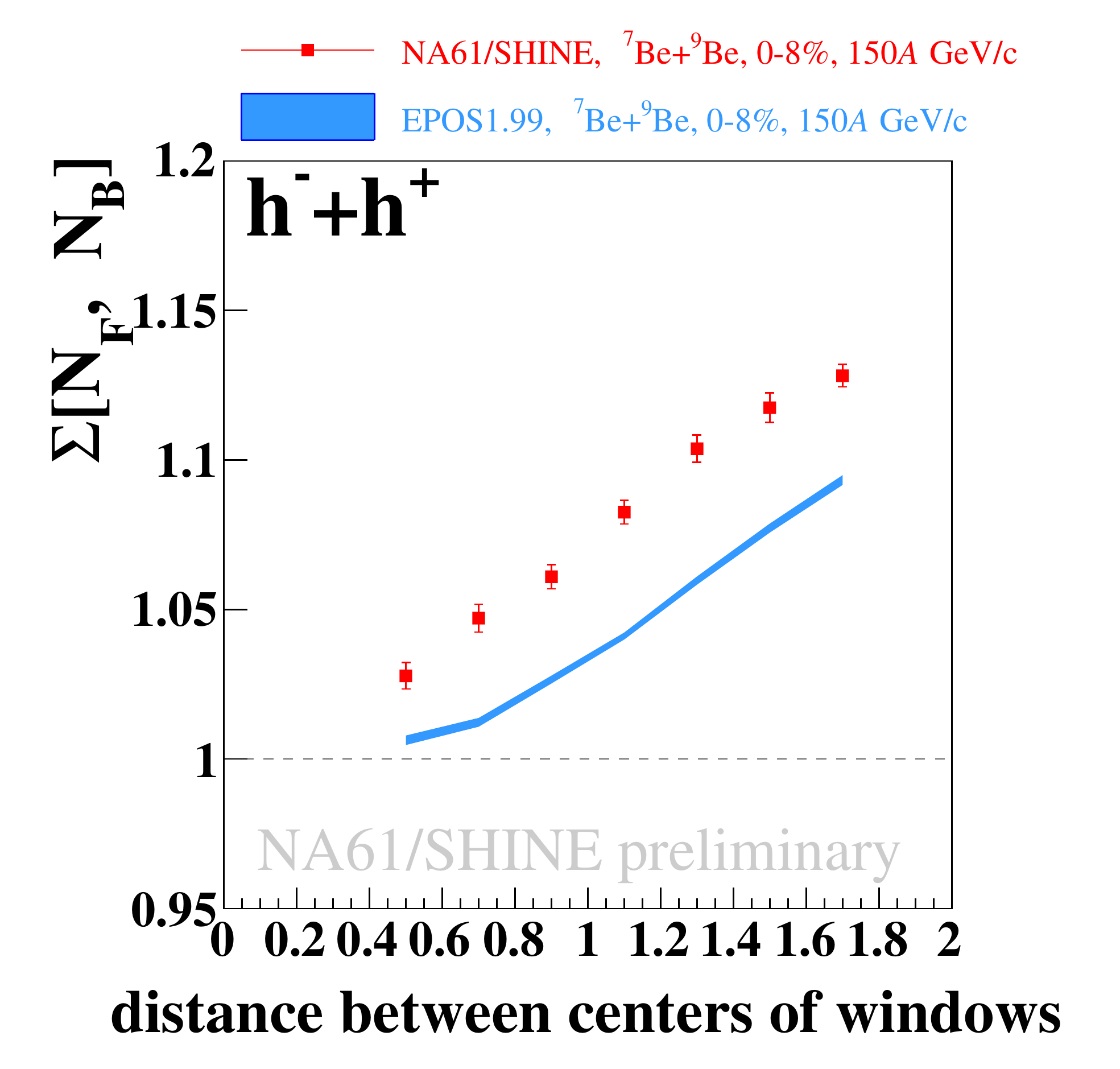} \begin{center}(a)\end{center}
\end{minipage}
\begin{minipage}{12pc}
\includegraphics[width=12pc]{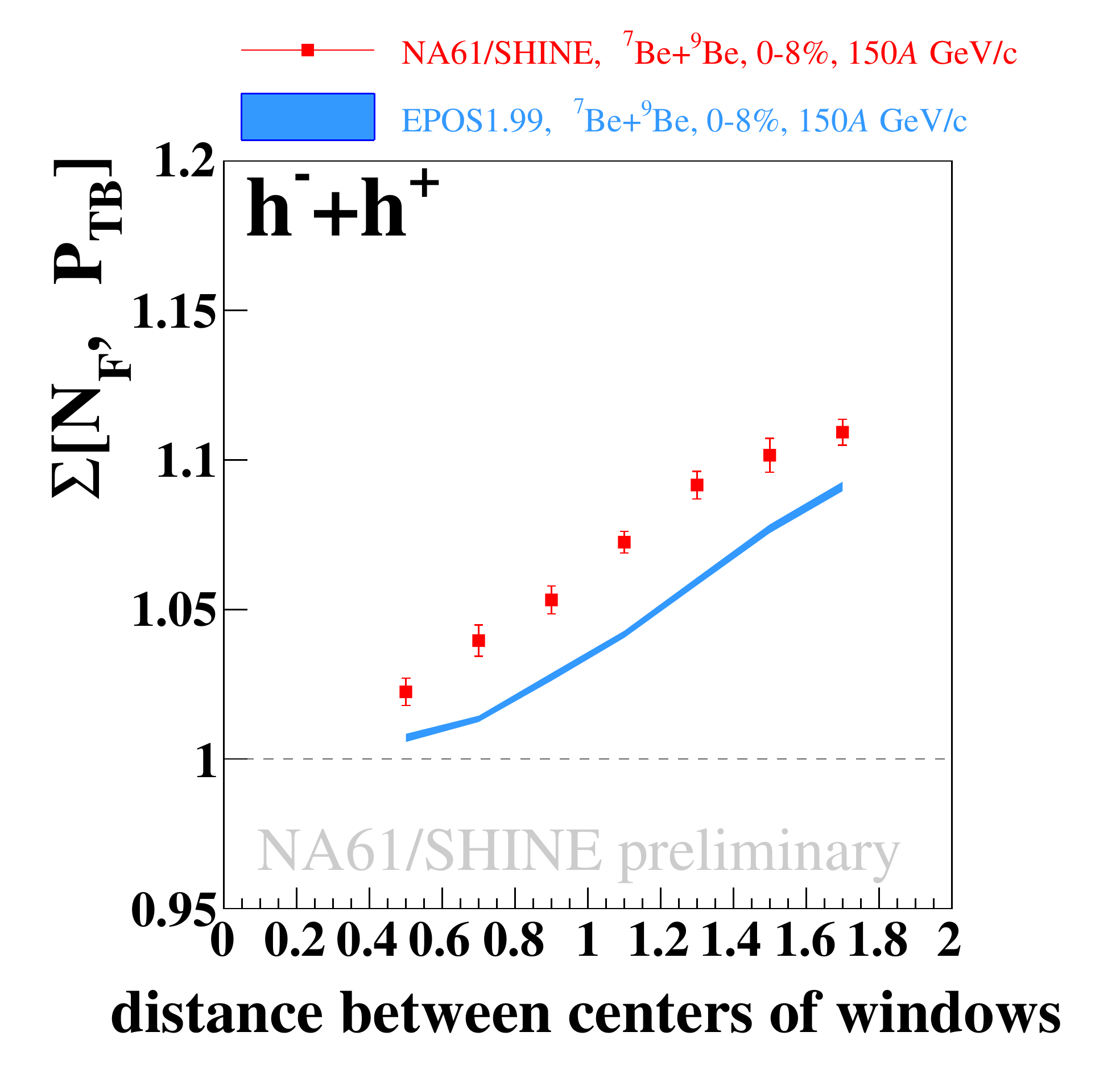} \begin{center}(b)\end{center}
\end{minipage}
\begin{minipage}{12pc}
\includegraphics[width=12pc]{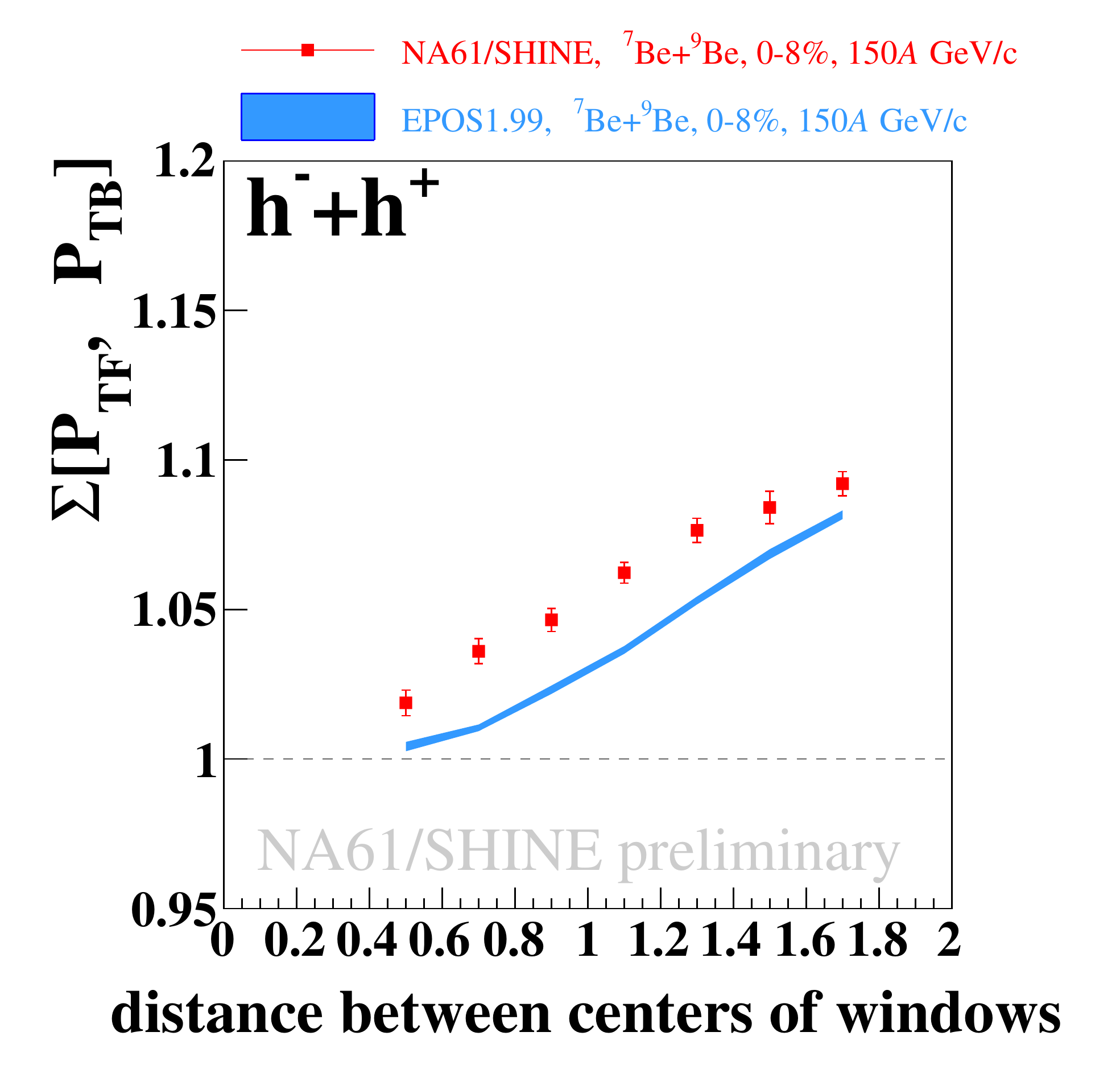} \begin{center}(c)\end{center}
\end{minipage}
\caption{\label{label3} Dependence of $\Sigma[\nF,\nB]$ (a), $\Sigma[\nF,\PB]$ (b) and $\Sigma[\PF,\PB]$ (c) on the separation between the two pseudorapidity windows. Red squares are preliminary NA61/SHINE measurements for 0-8$\%$ Be+Be collisions at 150{\it A} GeV/c. Blue band represents the EPOS1.99 model predictions.}
\end{figure}

The analysis was performed for 7 pairs of pseudorapidity intervals, with forward window fixed at (4.7,5.2) and the backward window moving from (3,3.5) to (4.2,4.7). Figure \ref{label3} shows preliminary results for the dependence of quantities (\ref{sigmaNBF})-(\ref{sigmaPTPTBF}) on the separation between the two windows. It is peculiar that all three quantities show qualitatively the same behaviour which is quite well reproduced by the EPOS1.99 model. Moreover, analogous behaviour was predicted by the quark-gluon string model \cite{Vechernin} for p+p interactions at LHC energies. The PHOBOS collaboration studied joint fluctuations of multiplicities in two windows quantified by the variance of $C=\frac{N_{F}-N_{B}}{\sqrt{N_{F}+N_{B}}}$ \cite{PHOBOS} which has similar properties like $\Sigma[\nF,\nB]$, although it is not strongly intensive. The variance of $C$ showed a similar trend with separation between rapidity windows in Au+Au collisions at $\sqrt{s_{NN}}=200$ GeV.

Similarly to the results in the previous section, no corrections for detector inefficiencies and trigger biases were applied as simulations have shown that their effect does not exceed 5$\%$.

The statistical uncertainties were estimated using the sub-sample method, and the systematic uncertainties were estimated to be smaller than 10$\%$.

\section{Conclusions} 
Dependence of the quantities $\Delta[P_{T},N]$ and $\Sigma[P_{T},N]$ on the width of the pseudorapidity interval were measured by the NA61/SHINE experiment in forward energy selected Be+Be collisions at 150{\it A} GeV/c. No indications of the critical point of strongly interacting matter were observed. However, a significant qualitative discrepancy between the experimental results on $\Delta[P_{T},N]$ and the EPOS1.99 model predictions is clearly visible.

NA61/SHINE preliminary results for $\Sigma[\nF,\nB]$, $\Sigma[\nF,\PB]$ and $\Sigma[\PF,\PB]$ on the separation of the forward and backward pseudorapidity intervals were also presented. It is the first measurement of this type at SPS energies. Experimental results are in good agreement with the EPOS1.99 model predictions. The observed trend to increase with separation is also similar to the predictions of the quark-gluon string model for LHC energies.  
 
\ack
This work was supported by the Russian Science Foundation under grant 17-72-20045.

\end{document}